\documentclass[journal]{IEEEtran}

\ifCLASSINFOpdf
\else
\fi
\usepackage{amsmath}
\usepackage{graphicx}
\usepackage{graphicx}
\usepackage{xcolor}
\usepackage[bottom]{footmisc}
\usepackage{epstopdf}
\usepackage[left=0.45in,right=0.45in,top=0.5in,bottom=0.45in, textheight=18in]{geometry}
\usepackage{algorithm}
\usepackage[noend]{algpseudocode}

\makeatletter
\def\BState{\State\hskip-\ALG@thistlm}
\makeatother

\begin{document}
\raggedbottom
%
\title{UAV-Assisted Heterogeneous Networks for
Capacity Enhancement}
%
%
%

\author{Vishal Sharma, 
        Mehdi Bennis, 
        Rajesh Kumar
\thanks{V. Sharma and R. Kumar are with Computer Science and Engineering Department, Thapar University, Patiala,
Punjab-147004, INDIA, e-mail: vishal\_sharma2012@hotmail.com, rajeshkumar\_nagdev@hotmail.com.}
\thanks{M. Bennis is with CWC - Centre for Wireless Communications, University of Oulu, Finland, Email: bennis@ee.oulu.fi.}

{\thanks{Manuscript received .....; revised ........}}}

\maketitle

\begin{abstract}
Modern day wireless networks have tremendously evolved driven by a sharp increase in user demands, continuously requesting more data and services. This puts significant strain on infrastructure based macro cellular networks due to the inefficiency in handling these traffic demands, cost effectively. A viable solution is the use of unmanned aerial vehicles (UAVs) as intermediate aerial nodes between the macro and small cell tiers for improving coverage and boosting capacity. This letter investigates the problem of user demand based UAV assignment over geographical areas subject to high traffic demands. A neural based cost function approach is formulated in which UAVs are matched to a particular geographical area. It is shown that leveraging multiple UAVs not only provides long range connectivity but also better load balancing and traffic offload. Simulation study demonstrate that the proposed approach yields significant improvements in terms of 5th percentile spectral efficiency up to 38\% and reduced delays up to 37.5\% compared to a ground-based network baseline without UAVs.
\end{abstract}


%
\IEEEpeerreviewmaketitle

\section{Introduction}
In order to tame the increasing data demands, small cell deployments are of key importance. To date, primary focus of small cell networks is to enhance the overall capacity by bringing users closer to serving base stations~\cite{li2014throughput}. Small cells can be deployed as standalone self organizing networks, or operating in conjunction with the existing macro cellular network. Due to the ever-increasing data traffic demands, an ultra-dense deployment of small cells is not a cost-effective strategy due to CAPEX/OPEX issues. At the same time, recent developments in unmanned aerial vehicles (UAVs), driven by Google and Facebook bring forward the idea of using UAVs for coverage extension and capacity enhancements. UAVs can be used as aerial access points acting as \emph{pivot} between macro and small cell tiers. UAVs can autonomously provide a reliable multi-connectivity in areas prone to high demand or link failures. UAVs can be used as aerial access points, or relays between disconnected networks and enhanced connectivity~\cite{guo2014performance}. A smart combination of all these networks can provide a vast range of applications in civilian networks~\cite{merwaday2015uav}. One of the major issues faced by these networks is on-demand/on-the-fly capacity provisioning. Capacity refers to data rate transmission towards ground users, whereas delays refer to latency of data transmission~\cite{mozaffari2015unmanned}. Using drone small cells as aerial support to existing cellular network can handle these high traffic situations more cost-efficiently. While deploying a single UAV is relatively easy using a maximum coverage point over the demand area, deploying more UAVs operating in coordination is more challenging due to interference from other aerial nodes~\cite{mozaffari2015drone}. Thus, an efficient approach is required that not only provides efficient topology for UAVs based on user demands, but also improved connectivity and enhanced coverage. Using multiple UAVs as relays between the existing macro cells and small cell networks is the primary focus of this letter, in which the goal is to increase capacity, and lower transmission delays.

In this letter, a cost function based multiple UAVs deployment model is presented. The proposed model uses user demand patterns to assign a cost and density function to each area and UAVs. These cost and density functions are then used to match each UAV to a particular demand zone via a reverse neural model based on user demand patterns~\cite{chandrashekarappa2014forward}.
\begin{figure}[!hb]
  \centering
\includegraphics[width=200px]{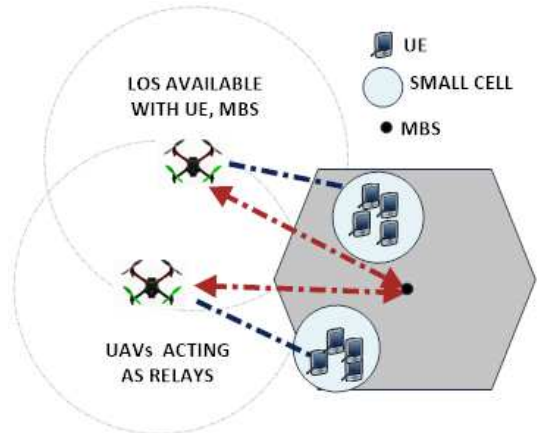}\\
  \caption{Illustration of an UAV-overlaid network deployment, with UAVs acting as relays between MBS and UEs.}\label{fig1}
\end{figure}
\section{System Model}
\begin{figure}[!hb]
  \centering
\includegraphics[width=200px,height=180px]{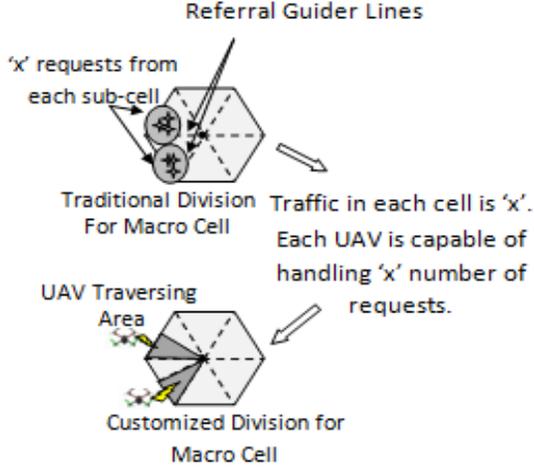}\\
  \caption{Area distribution in the macro cell with an illustration of referral guider lines.}\label{fig3}
\end{figure}

The proposed model aims at provisioning continuous data between macro and small cell user equipments (UEs). UAVs further enhance load balancing by forming multiple intermediate links between the macro cell and the small cell UEs. The proposed model focuses on UAV-to-UE links rather the MBS-to-UAV backhaul capacity links. A traditional small and macro cell network is shown in Fig.~\ref{fig1}. This network provides a direct link between the UEs and the macro cell base station (MBS). However, with a continuous increase in number of users within the coverage of a particular small cell, it becomes almost impossible to sustain connectivity without any loss of data. This is a problematic issue for future generation 5G networks, whereby demand is set to be greater than the available capacity. This is the focus of this work harnessing the formation of UAVs for a reliable and load balanced network. The proposed approach leverages a cost based framework which uses neural demand patterns to identify areas with high demands. A predictive chart is formed at the MBS acting as a launch point for UAVs. This chart helps in finding appropriate positions and topology for UAVs. Network stability is attained by minimizing a cost function associated with both demand areas and deployed UAVs. A delay threshold is defined to analyze the network performance, defined as the network state with minimum delay in providing capacity coverage in high demand areas, minimum errors in mapping UAVs to particular demand area, and high reliability in terms of data rate delivery. The model consists of deploying $n$ UAVs each one capable of handling $S_{n}$ service requests in a zone governed by an MBS. If $S_{r}$ requests are made by active users in a small cell zone that the MBS is unable to handle, then the minimum number of UAVs required to handle this traffic is $\frac{S_{r}}{S_{n}}$. Moreover, the optimal placement of multiple-UAVs in the required zone is a major issue. For this, the concept of \emph{zone guider lines} is considered where \emph{zone guider lines} divide a particular area into a set of small regular areas acting independently. For a low-complexity solution, the number of UAV is kept equal to the number of guider lines. The traditional hexagonal cell is divided into standard guider lines, then, the area with high user requests are marked. Next, the existing guider lines form the maximum and minimum limit for introducing new guider lines which will embark the area to be governed by a UAV. This procedure is presented in Fig~\ref{fig3}.
\subsection{Assumptions}
In order to validate the effectiveness of the proposed approach, some generic assumptions are as follows:
\begin{itemize}
  \item UAVs operate on same frequency spectrum.
  \item Each UAV is of the same make, whose configuration does not affect its positioning.
\end{itemize}
\section{Proposed Approach}
UAVs are used as high altitude base stations to cover certain geographical areas. The proposed approach uses a cost based neural model to find the appropriate user demand zones where UAVs is placed. The cost function includes the cost of operation which has to be minimum, and the cost of handling UAVs. Thus, the proposed approach focuses on finding the appropriate cost function for demand areas and UAVs, allocating them to MBS, and defining a neural model to minimize the cost function. In the proposed model, UAVs operate at an altitude $h$ over an area $A$, with $x$ number of users. Users service request $S_{r}$ come with an arrival rate of $\gamma$, and a mean packet size of 1/$\mu$. The load/delay, denoted by $L$ (in seconds) for a user at location $y$, is computed as~\cite{samarakoon2014opportunistic}:
\begin{equation}\label{eq:d1}
L (y) =\frac{\gamma}{W \;\log(1+SINR(y))\times \mu}.
\end{equation}
The channel modeling includes radio range and pathloss. Moreover, round robin approach is applied for scheduling. Further, the system model is defined with respect to UAVs rather than the MBS, assumed to operate on orthogonal band. The area load $L_{a}$ is given by~\cite{samarakoon2014opportunistic}:
\begin{equation}\label{eq:d2}
L_{a}=\int_{y \in A} L(y) dy.
\end{equation}
Here, $W$ is the system bandwidth, and assuming that UAVs operate on the same frequency spectrum, the signal-to-interference-plus-noise ratio $\left(SINR\right)$ from the $i^{th}$ UAV to a given UE at location $y$, considering UAV-to-UAV interference, is:
\begin{equation}
SINR(y)= \frac{\frac{P \; K  }{R_{iy}^{\alpha}}}{\sum_{j=1, j\neq i}^{n} \frac{P \;K}{ R_{jy}^{\alpha}} + N_{0}},
\end{equation}
where $P$ is the UAV transmission power, $K$ is a factor that accounts for the geometrical parameters such as transmitter and receiver antenna heights, $R_{iy}$ is the distance between the $i^{th}$ UAV and the UE at location $y$, $\alpha$ is the path loss exponent, and $N_{0}$ is the noise power spectral density. The spectral efficiency $N_{S}$ for a user at location $y$ following round robin scheduling is given by:
\begin{equation}
N_{S}=W.\;\frac{\log_{2} \left( 1+SINR(y) \right)}{x}
\end{equation}
The cost function is a function of capacity, delay, availability of line of sight (LOS), and coverage. Further, let $D_{f}$ denote the density function that quantifies the population of active/non-active users based on users' request patterns. $D_{f}$ accounts for the number of active users $x$, packet loss (call drops) $C_{d}$, service requests $S_{r}$, and the total amount of users a cell can handle is $T_{r}$. For the considered network, two variants of the density function, one for a given area $D_{f}^{A}$, and the other for UAVs $D_{f}^{U}$ are computed as
\begin{equation}\label{eq:1}
D_{f}^{A}=\min\left(\frac{\left(\frac{x}{T_{r}}\right)^{S_{r}}\;e^{- \left(\frac{x}{T_{r}}\right)}}{S_{r}!}\right),
\end{equation}
and
\begin{equation}\label{eq:2}
D_{f}^{U}=\min\left(\frac{\left(\frac{L_{a}}{n}\right)^{S_{n}}\;e^{- \left(\frac{L_{a}}{n}\right)}}{S_{n}!}\right),
\end{equation}
respectively. $D_{f}^{A}$ accounts for user distribution over the area, in which a higher value requires more UAVs, and a minimum value shows the efficient connectivity with no further requirement of intermediate relays. $D_{f}^{U}$ accounts for pending user requests in area A with respect to the number of UAVs deployed. A higher value denotes the requirement of more UAVs, and a minimum denotes efficient service handling using current deployment. Here, $\frac{x}{T_{r}}$ denotes the ratio of active users to the total users a cell can handle. For 100\% accuracy in mapping, $\frac{x}{T_{r}}=1$. (\ref{eq:1})-(\ref{eq:2}) account for the cost function provided constraint (\ref{eq:3}) holds.
\begin{equation}\label{eq:3}
\sqrt{\frac{1}{S_{r}} \sum_{i=1}^{S_{r}} \left( T_{r}^{i}-C_{d}^{i}\right)} \leq \frac{x}{T_{r}}.
\end{equation}
For an efficient operation, the deviation in (\ref{eq:3}), the number of users with unhandled service requests, should be kept minimum. Furthermore, the per area and UAV cost function $C_{f}^{A}$ and $C_{f}^{U}$ is given as:
\begin{equation}\label{eq:4}
C_{f}^{A}=\min\left(D_{f}^{A}\; L_{a}\; \left(\eta_{1}S_{r} + \eta_{2}T_{r} \right)\right),
\end{equation}
and
\begin{equation}\label{eq:5}
C_{f}^{U}=\min\left(D_{f}^{U}\; R^{\alpha}_{iy}\; \left( \eta_{1}S_{r}+\eta_{2} x\right)\right), LOS=true,
\end{equation}
respectively. Here, $\eta_{1}$ and $\eta_{2}$ are network balancing constants such that $\left(\eta_{1}, \eta_{2} \right) \in \left(0,1\right)$. In general, $\eta_{1}$ is driven by the network bandwidth and link speed, whereas $\eta_{2}$ is driven by the number of active connections. For ideal state, $\eta_{1}$ and $\eta_{2}$ equals 1. In general, $0.5 \leq \eta_{1} \leq 1$ and $\eta_{1} \leq \eta_{2} \leq 1$ which denotes that the network transfer rate must be higher than half of the initial configured rate. Both cost functions $C_{f}^{A}$ and $C_{f}^{U}$ are governed by the constraints of $D_{f}^{A}$ and $D_{f}^{U}$. The complete availability of LOS is one of the key driving factor for continuous connectivity. The overall cost function $C_{f}^{O}$ is computed at the MBS to maintain the overall network connectivity such that
\begin{equation}\label{eq:6}
C_{f}^{O}=\min\left( \frac{1}{n} \sum_{i=1}^{n} \left(C_{f}^{U}\right)_{i} + \sum_{j=1}^{A_{T}} \left(\frac{C_{f}^{A}}{U_{T}} \right)_{j}\right).
\end{equation}
Here, $U_{T}$ is the number of UAVs allocated to a particular area, $A_{T}$ is the number of total demand areas. With more UAVs, more resources are available in terms of transmission power, yielding high throughput and reduced delay. Further, the delay ($L_{d}$) at each node is computed as:
\begin{equation}
L_{d}=L_{transmission}+L_{propagation}+L_{queue}+L_{processing}.
\end{equation}
Here, $L_{transmission}$ is the transmission delay defined as the load/delay of a particular user and is equal to $L$ (\ref{eq:d1}), $L_{propagation}$ is the ratio of the distance between nodes to the propagation speed, $L_{queue}$ is the waiting time of the packet, and $L_{processing}$ is the network operational time.
\section{Neural Demand Patterns and Network Capacity}
The goal is to optimize the density functions and minimize the cost functions defined in (\ref{eq:1}), (\ref{eq:2}), (\ref{eq:4}), (\ref{eq:5}), and (\ref{eq:6}). By controlling $D_{f}^{A}$ and $D_{f}^{U}$, the user distribution with pending requests are controlled, which in turn, minimizes $C_{f}^{A}$, $C_{f}^{U}$, and $C_{f}^{O}$. This minimization provides guaranteed service to UEs. Thus, the aim of the neural model is to accurately map UAVs to demand areas so as to minimize these cost functions.
\begin{figure}[!ht]
  \centering
\includegraphics[width=180px,height=170px]{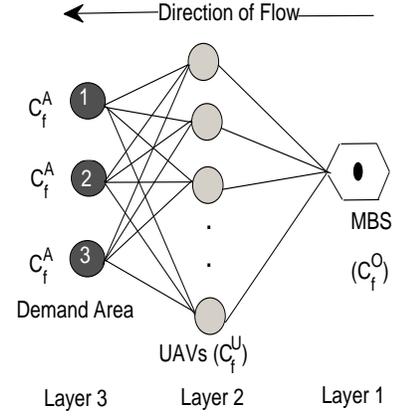}\\
  \caption{Reverse neural model modeling the user demand pattern: $C_{f}^{A}$ refers to area cost function, $C_{f}^{U}$ refers to UAV cost function, $C_{f}^{O}$ is the overall cost function.}\label{fig4}
\end{figure}
Demand patterns are used to minimize the demand and cost functions by efficiently deploying aerial nodes as intermediate nodes between high demand areas and the MBS. These demand patterns are driven by a reverse multi-hierarchical neural model which is a combination of input, hidden, and output layer. Although neural models are slow and complex in operation, the reverse neural model accounts for accurate mapping of UAVs to demand areas with lesser iterations. This provides a low-complexity approach for UAV-to-Area mapping. The output for the proposed model is computed in forward direction, i.e. from Layer 3 to Layer 1, but nodes are deployed in reverse direction i.e. from Layer 1 to Layer 3, as shown in Fig.~\ref{fig4}. Input involves demand areas that require services of UAVs. MBS is the driving factor of this reverse neural model, and it act as an output layer. UAVs are the intermediates, and are not having permanent connections with the demand areas, thus, acts as a hidden layer. These neural patterns are then topologically rearranged to form a stable network with minimized cost function. Layer 3 defines the cost function for underlying demand areas, layer 2 is for aerial nodes, and output layer 1 is mapped to the MBS. At first, a mesh like interface is initialized for the neural model (Fig.~\ref{fig4}), then, the whole model is rearranged to allocate links that will lower the cost functions of all the zones. This procedure is governed by series of steps given in Algorithm~\ref{algo1}.
\begin{algorithm}[!ht]
\fontsize{7}{9}\selectfont
\caption{UAV to Area Mapping}
\label{algo1}
\begin{algorithmic}[1]
\State \textbf{Input}: $U \longleftarrow UAVs$, $A \longleftarrow Demand \;areas$
\State Initialize Network
\State Divide A into subdivisions $D_{M}$
\While{(U are not mapped to subarea $D_{M}$ \&\& ($C_{f}^{O}$=minimum))}
\State Store $D_{M}$ in area[] following descending order for $C_{f}^{A}$
\State i=1
\While{($i \leq count(area[])$)}
\State allocate U to $D_{M}$, such that min($C_{f}^{U}$) is mapped to max($C_{f}^{A}$)
\State Compute $C_{f}^{A}$, $C_{f}^{U}$, $C_{f}^{O}$
\If {((U is mapped to $area[i]$) \&\& ($C_{f}^{A}(i)$=minimum))}
\State continue
\Else
\State re-initialize $U$, $C_{f}^{A}$
\State reset
\EndIf
\State \textbf{end if}
\State $i=i+1$
\EndWhile
\State \textbf{end while}
\EndWhile
\State \textbf{end while}
\end{algorithmic}
\end{algorithm}
Initially, the demand area $A$ is subdivided into small segments $D_{M}$, and the cost function is computed for each of the subdivided area. Each of the cost function is ranked in descending order, in which the UAV with a minimum cost function is mapped to the area with maximum cost function. This helps in balancing the overall load of the network. After initial allocation, all the cost functions are recomputed, and UAVs are mapped to the next high demand area based on the cost function. This procedure continues till a minimum is attained. With an efficient deployment, the cost function is minimized by sub-dividing the density function based on the area so that active users becomes equal to the total registered users i.e. $\frac{x}{T_{r}}=1$ such that equation (\ref{eq:1}) reduces to
\begin{equation}
D_{f,avg}^{A}=\frac{1}{e\times S_{r}!}
\end{equation}
where e=2.71828 approx. For verification at any stage, the average density function of each area must be less than equal to $D_{f,avg}^{A}$.
\begin{table}[!ht]
\fontsize{7}{9}\selectfont
\centering
\caption{Parameter Configurations}\label{self_conf}%
\begin{tabular}{l l l}
\hline\\
\textbf{Parameter} & \textbf{Value} & \textbf{Description}\\
\hline
\hline\\
$A$ & 10000x10000 sq. m. & Simulation Area\\
MBS & 10& Number of Macro Cell Base Station\\
$T_{r}$ & 1200 (per MBS) & Max Users in a Cell\\
$n$& 6 (per MBS)& Number of UAVs\\
$S_{n}$ & 200 & Service Requests handled by each UAV\\
$N_{0}$ & -170 dBm/Hz & Noise Power Spectral Density\\
$\frac{1}{\mu}$ & 1024 B&Packet Size \\
$h$ &200-500 Feet& UAV altitude\\
$\frac{\gamma}{\mu}$ & 256 kbps& Offered Traffic\\
$\alpha$& 4&Path loss Exponent\\
$K$ & -11 dB & Transmission Constant\\
$P$ & 35 dBm & UAV Transmission Power \\
$S_{r}$& 30-50 per zone& Service Requests\\
$W$ & 10 MHz& System Bandwdith\\
$x$ & 400 & Active Users\\
\hline
\end{tabular}
\end{table}%
\begin{figure}[!ht]
\begin{minipage}[t]{0.20\textwidth}
\centering
\includegraphics[width=120px,height=80px]{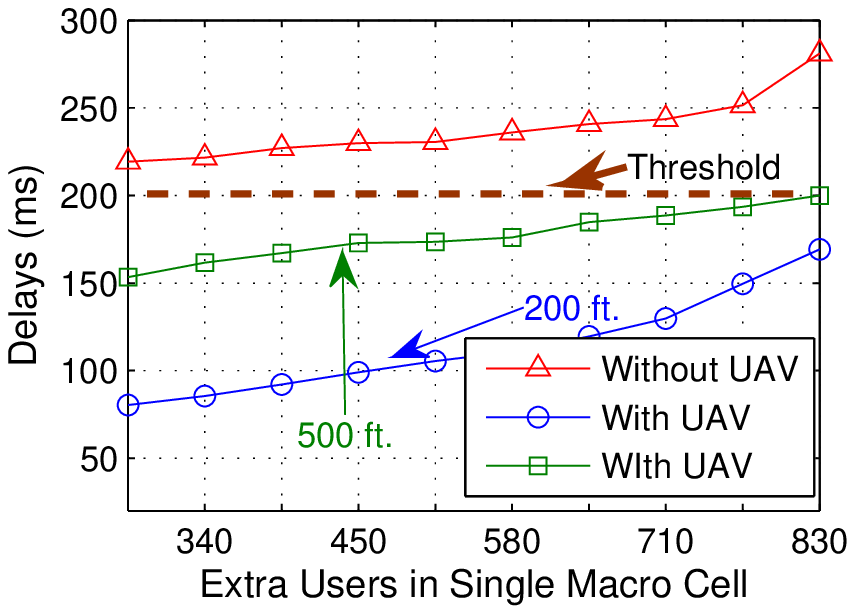}
\caption{\fontsize{6}{6}\selectfont Networks Delays vs. Extra Users }
\label{g1}
\end{minipage}
\hspace{\fill}
\begin{minipage}[t]{0.26\textwidth}
\centering
\includegraphics[width=120px,height=80px]{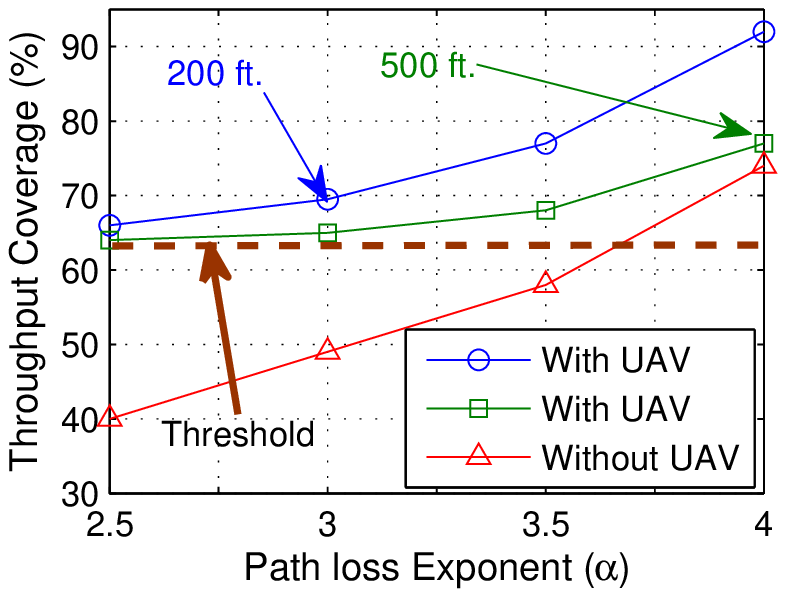}
\caption{\fontsize{6}{6}\selectfont Throughput Coverage vs. Path loss Exponent}
\label{g3}
\end{minipage}
\end{figure}
\begin{figure}[!ht]
\begin{minipage}[t]{0.20\textwidth}
\centering
\includegraphics[width=120px,height=80px]{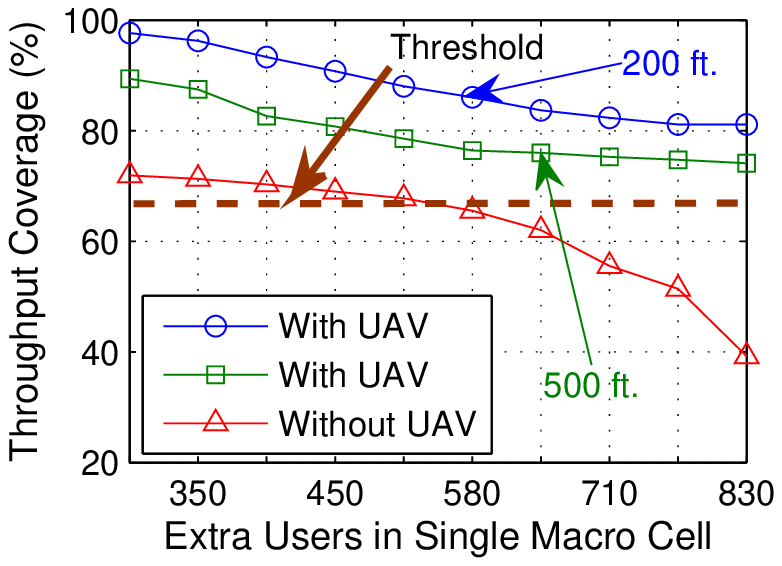}
\caption{\fontsize{6}{6}\selectfont Throughput Coverage vs. Extra Users}
\label{g4}
\end{minipage}
\hspace{\fill}
\begin{minipage}[t]{0.26\textwidth}
\centering
\includegraphics[width=120px,height=80px]{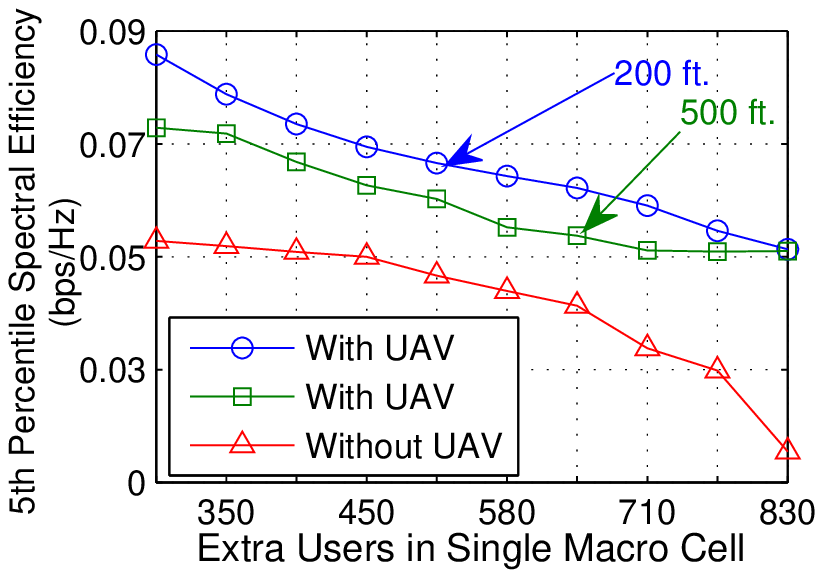}
\caption{\fontsize{6}{6}\selectfont 5th Percentile Spectral Efficiency vs. Extra Users }
\label{g5}
\end{minipage}
\end{figure}
\begin{figure}[!ht]
\begin{minipage}[t]{0.20\textwidth}
\centering
\includegraphics[width=120px,height=80px]{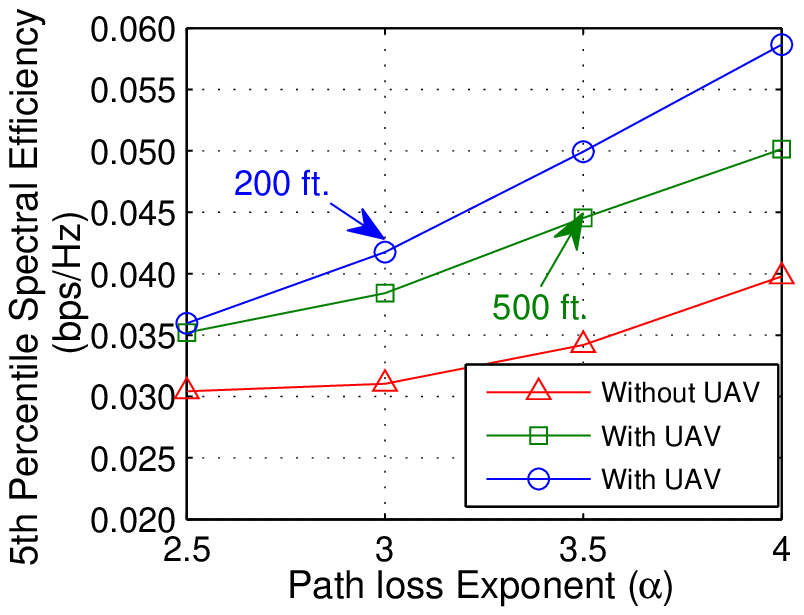}
\caption{\fontsize{6}{6}\selectfont 5th Percentile Spectral Efficiency vs. Path loss Exponent }
\label{g6}
\end{minipage}
\hspace{\fill}
\begin{minipage}[t]{0.26\textwidth}
\centering
\includegraphics[width=120px,height=80px]{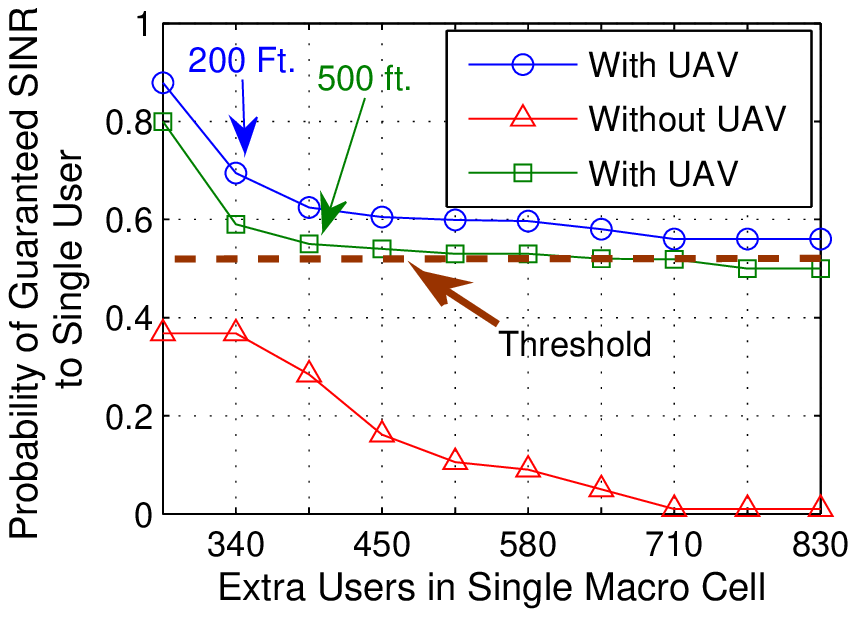}
\caption{\fontsize{6}{6}\selectfont Probability of Guaranteed SINR vs. Extra Users }
\label{g7}
\end{minipage}
\end{figure}
\section{Performance Evaluation}
The proposed model is analyzed using network simulations, to efficiently allocates areas with higher demand patterns to UAVs based on their cost function. All parameters and configurations are presented in Table~\ref{self_conf}. Active users refer to the number of users in a cell requesting services. Thus, at any time instance, more than the maximum supported user requests can be present in the cell. Results were recorded for delays, network capacity, reliability, and value of cost function with and without use of UAVs. Delays were traced for complete data sharing with 1000 iterations of user requests over 1000 seconds. Availability of LOS is the primary condition for UAVs to form a link with the UE. Finding appropriate position in the user demand areas causes UAVs to adjust their altitude to guarantee LOS towards the UE. High altitude provides less interference and appropriate LOS, but also induces more delays. Thus, an optimum altitude with availability of LOS is required for better coverage. For analysis, the altitude was varied between 200 ft. and 500 ft. with a multi-antenna relay support for communication and backhaul link capacity of 1.2 Gbps. Delay threshold is fixed at 200 ms, which defines the upper limit above which the packet drop increases abruptly. Results show that the proposed approach leveraging UAVs yields 37.7\% lesser delays in comparison with a network comprising of small cell and macro cell. The performance delay for various UAV altitudes is given in Fig.~\ref{g1}. Throughput coverage defined as the percentage of users whose SINR is above the threshold (0.03 bps/Hz) is shown in Fig.~\ref{g3}. Therein, the use of UAVs increases the overall 5th percentile throughput coverage by 15.5\%, as shown in Fig.~\ref{g3} and Fig.~\ref{g4}. Moreover the optimal placement of UAVs according to user demand leveraging the reverse neural deployment model enhances the 5th percentile spectral efficiency. The accurate mapping optimally places UAVs according to demand patterns, thus, improving the 5th percentile spectral efficiency by 38\% approx., as shown in Fig.~\ref{g5} and Fig.~\ref{g6}. Finally, Fig.~\ref{g7} plots the probability of guaranteed SINR for a particular user in a given macro cell. Clearly, it can be noticed that the use of UAVs provides much guaranteed SINR above the threshold defined by $\eta_{1}$ in (\ref{eq:5}). Here, the SINR threshold is kept at 0.55 defining the value below which the network is unable to provide efficient connectivity to users. Results show that the proposed user demand based network model is capable of providing better capacity and prolonged connectivity than the existing cellular network.
\section{Conclusion}
In this letter, user demand based network model is proposed using multiple UAVs. The proposed model uses density and cost functions to compute areas with higher demands, whereby UAVs are deployed based on these cost functions. Analysis proved that the proposed model is capable of providing better capacity, reliability, and prolonged connectivity in comparison to existing ground-based wireless networks.
\bibliographystyle{ieeetr}
\nocite{*}
\bibliography{final_ref}








\end{document}